\documentclass[aps,prl,amssymb,superscriptaddress,nofootinbib,twocolumn,preprintnumbers]{revtex4-2}
\usepackage[utf8]{inputenc}
\usepackage[english]{babel}
\usepackage{amsmath}
\usepackage{amsfonts}
\usepackage{amssymb}
\usepackage{amsthm}
\usepackage{color}
\usepackage{graphicx}
\usepackage[force]{feynmp-auto}
\usepackage{feynmp}
\usepackage[hidelinks]{hyperref}

\newcommand\bw{\begin{widetext}}
\newcommand\ew{\end{widetext}}

 \def\be{\begin{equation}}
\def\ee{\end{equation}}
 \def\ba{\begin{align}}
\def\ea{\end{align}}
\def\bea{\begin{eqnarray}}
\def\eea{\end{eqnarray}}

\def\a{\alpha}
\def\b{\beta}

\def\m{\mu}
\def\n{\nu}

\begin{document}
\title{{\bf Time Orientability and Particle Production from Universal Horizons}}

\author{Francesco Del Porro}
\email[]{fdelporr@sissa.it}

\author{Mario Herrero-Valea}
\email[]{mherrero@sissa.it}

\author{Stefano Liberati}
\email[]{liberati@sissa.it}

\author{Marc Schneider}
\email[]{mschneid@sissa.it}

\address{SISSA, Via Bonomea 265, 34136 Trieste, Italy and INFN Sezione di Trieste}
\address{IFPU - Institute for Fundamental Physics of the Universe \\Via Beirut 2, 34014 Trieste, Italy}

\begin{abstract}
We discuss particle production in spacetimes endowed with a universal horizon in Einstein--Aether and Ho\v rava gravity. We argue that continuity and differentiability of the lapse function require the orientation of the foliation in the interior of the horizon to be reversed with respect to the exterior one. Unless this is allowed, interaction of gravitating scalar fields with the universal horizon leads to unitarity violations in the quantum theory. This property is responsible for particle production by the universal horizon, as we show by computing explicitly its Hawking temperature for all stationary and spherically symmetric spacetimes. We particularize our result to known analytic solutions, including those compatible with observational constraints.
\end{abstract}

\maketitle
\section{Introduction}
The prediction that black holes radiate is one of the few glimpses on Quantum Gravity available so far. Furthermore, the finding that Hawking's radiation is thermal points down to a profound connection between gravity and thermodynamics. Black holes in General Relativity (GR) have entropy, hinting to the existence of microscopic degrees of freedom whose collective dynamics gives rise to the gravitational interaction. In GR, all these properties are controlled by the horizon of the black hole, which acts as a causal barrier for information \cite{Jacobson:2003vx}.

This vision is challenged in models of Lorentz violating gravity, such as Einstein--Aether 
(EA) gravity \cite{Jacobson:2000xp}, where a time-like vector $U^\m$, the aether, breaks boost invariance. It defines a preferred time direction, and allows to write operators with only higher derivatives along spatial directions in the preferred frame, by a direct coupling to the projector orthogonal to $U^\m$. This avoids the introduction of ghost degrees of freedom, and allows for modified dispersion relations of the generic form 
\begin{align}\label{eq:disp_relation}
    \omega^2=q^2+\frac{\a_4 q^4}{\Lambda^{2}}+\dots +\frac{\a_{2n} q^{2n}}{\Lambda^{2n-2}},
\end{align}
which can propagate faster than light. Here $\Lambda$ is the scale of Lorentz violations and $\alpha_i$ are dimensionless constants. This questions the universality of the thermodynamic picture of black holes, since no universal causal boundary seems possible for all propagating speeds.

The intuitive image that we just described is however wrong in the case of known non-rotating black hole solutions in EA gravity \cite{Berglund:2012bu,Barausse:2012qh,Barausse:2015frm}. For those, $U^\m$ is irrotational and defines a preferred foliation onto space-like hypersurfaces, described by a scalar field $\Theta$
\begin{align}
    U_\m=\frac{\partial_\m \Theta}{\sqrt{|\partial_\a \Theta \partial^\a \Theta|}}.
\end{align}
When this happens, the dynamics of the theory coincides with the low energy limit of Ho\v rava Gravity \cite{Horava:2009uw}, known as Khronometric theory \cite{Blas:2010hb}.

In certain static space-times, it might happen that a foliation leaf becomes a constant radius hypersurface. Since all signals must travel by following the evolution of the subsequent leafs, this implies that no signal can escape from the region enclosed. Such a hypersurface is named \emph{universal horizon} (UH), and it occurs when $(\chi\cdot U)=0$, where $\chi^\m$ is the Killing vector defining staticity. The UH defines an absolute causal boundary for all signals, regardless of their propagation speed. Due to this, one may be tempted to generalize the laws of black hole thermodynamics to the UH. However, it is not clear if this naive approach is correct. Although there are works that point towards this direction \cite{Blas:2011ni,Berglund:2012fk,Pacilio:2017swi,Ding:2015fyx,Ding:2016srk,Cropp:2016gkn,Herrero-Valea:2020fqa}, there are others that question this picture \cite{Michel:2015rsa}. Among the positive results, the resulting Hawking temperature reads
\begin{align}\label{eq:Hawking_temp}
    T=\frac{n-1}{n}\frac{\kappa_{\text {\sc uh}}}{\pi},
\end{align}
for a dispersion relation in the ultra-violet (UV) of the form $\omega^2\sim \alpha_{2n} q^{2n}$. Here $\kappa_{\text {\sc uh}}=-\left.\frac{1}{2}(a\cdot\chi)\right|_{r_{\text {\sc uh}}}$ is the surface gravity of the UH, and $a_\m=U^\a \nabla_\a U_\m$ is the acceleration of the aether. This jeopardizes the universality of thermodynamics, unless the UV behavior of the dispersion relation is universal\footnote{Universal dispersion relations arise for instance in Ho\v rava Gravity, induced by loop corrections \cite{Pospelov:2010mp}.}. Incidentally, it is curious that a necessary condition for the existence of a UH is that $\left. (a\cdot\chi)\right|_{\rm UH}\neq 0$ \cite{Bhattacharyya:2015gwa}. 

However, there is an issue that has been overlooked so far. If the foliation describing the interior of the UH is oriented in the same direction as the one in the exterior region, the surface gravity of the UH is ill-defined, showing a discontinuity. As a consequence, the lapse function will always exhibit a kink at the UH, unless we allow the interior and exterior leafs to \emph{not} be ordered in the same direction, thus breaking global causality in a very special, yet not worrisome, manner. 

In this \emph{paper} we show how, when both foliations are oriented in the same direction, the quantum theory for a scalar field \emph{in the exterior of the black hole} violates the axioms of probability close to the UH. When the reversal of the black hole interior is allowed instead, standard thermal properties are recovered. As a final result of our work, we derive the Hawking temperature of the radiation for all spherically symmetric and stationary metrics with a UH.

\section{Black Holes in EA gravity}
In the following we consider stationary and spherically symmetric solutions in $d=4$ dimensions, whose most general metric reads
\begin{align}\label{eq:metric}
ds^2=F(r)dt^2 -\frac{B(r)^2}{F(r)}dr^2  -r^2 d\Omega_2^2,  
\end{align}
where we use mostly minus signature and we assume asymptotic flatness. The corresponding aether is irrotational and has a single degree of freedom $A(r)$. We choose to normalize it as \cite{Berglund:2012bu}
\begin{align}\label{eq:aether}
    U_\m dx^\m=\frac{1+F(r)A(r)^2}{2 A(r)}dt +\frac{B(r)}{2A(r)}\left(\frac{1}{F(r)}-A(r)^2\right)dr,
\end{align}
where the form of the $U_r$ component is fixed by demanding $U_\m U^\m=1$. 

Staticity of the solution defines a Killing vector, which takes the form $\chi_\m dx^\m=F(r) dt$ and signals the position of a Killing horizon whenever $|\chi|^2=F(r)=0$. The sign of the aether components is chosen so that it is future directed in the asymptotic region.

The aether defines a preferred time direction in terms of a foliation in co-dimension one hypersurfaces $N d\tau =U_\m dx^\m$ where 
\begin{align}
    N=(\chi\cdot U)=\frac{1+F(r)A(r)^2}{2 A(r)},
\end{align}
is the lapse function of the foliation. This implies
\begin{align}
    dt=d\tau +\frac{B(r)}{F(r)}\left(\frac{F(r)A(r)^2-1}{F(r)A(r)^2+1}\right) dr.
\end{align}

In this preferred frame, the metric takes the Arnowitt–Deser–Misner (ADM) form\footnote{Hereinforward, we use Greek letters to refer to 4-dimensional indices, while Latin letters run only over spatial directions.} \cite{Arnowitt:1962hi}
\begin{align}\label{eq:ADM}
    ds^2 = N^2 d\tau^2 -\gamma_{ij}(dx^i + N^i d\tau)(dx^j + N^j d\tau),
\end{align}
where $N^i$ is the shift vector and $\gamma_{ij}$ the spatial metric. 

This choice of lapse is not unique. The theory is invariant under foliation preserving diffeomorphisms (FDiff)
\begin{align}\label{eq:FDiff}
    &\tau \rightarrow \tilde\tau(\tau), \qquad x^i\rightarrow \tilde x^i(\tau,x).
\end{align}
The different elements of the metric transform under this as
\begin{align}\label{eq:metric_transf_FDiff}
    &\tilde N = \frac{d\tau}{d\tilde \tau}N,\quad \tilde N ^i= \left(N^j\frac{\partial\tilde x^i}{\partial x^j}-\frac{\partial\tilde x^i}{\partial \tau}\right)\frac{d\tau}{d\tilde \tau},
\end{align}
while the spatial metric $\gamma_{ij}$ transforms as a tensor under spatial diffeomorphisms, but it is invariant under time reparametrizations. Note that by going to this frame, the time coordinate gets identified with the scalar field $\Theta=\tau$.

We introduce now a simple notion of causality\footnote{This corresponds to the notion of pre-causality discussed in \cite{Liberati:2013xla}. For a deeper discussion of causality in foliated space-times space-times cf. \cite{Bhattacharyya:2015gwa}.} by demanding that \emph{all clocks tick in the same direction}. By observing \eqref{eq:metric_transf_FDiff} we note that this requires
\begin{align}
    N>0,
\end{align}
for all foliation leafs and observers. Whenever this condition fails to be fulfilled, we can find an observer with acausal time evolution within the region where $N<0$, since there $d\tau/d\tilde\tau<0$.

The solution will have a UH whenever the following two conditions are satisfied \cite{Bhattacharyya:2015gwa}
\begin{align}
    \left.(\chi \cdot U)\right|_{r_{\text {\sc uh}}}=0,\qquad \left.(\chi\cdot a)\right|_{r_{\text {\sc uh}}}\neq 0.
\end{align}

Since $\chi^\m$ is a Killing vector and there is no explicit time dependence in any metric nor aether component, we have
\begin{align}\label{eq:adotchi}
  (\chi\cdot a)=U^\m \partial_\m (\chi\cdot U)=U^r \partial_r N,
\end{align}
so these conditions can be rewritten as
\begin{align}\label{eq:UH_cond_lapse}
    \left.N\right|_{r_{\text {\sc uh}}}=0,\quad \left.\partial_r N\right|_{r_{\text {\sc uh}}}\neq 0.
\end{align}

The fact that $N$ vanishes at the UH is not a problem. It simply implies that the time of the preferred observer corresponding to that particular lapse diverges, in order to keep $U^\m$ finite. This is just a coordinate singularity of \eqref{eq:ADM}. What cannot be avoided, however, is what we want to remark here. If we demand $U^\m$ and all metric invariants to be of class $\mathbf{C}^2$, so that the action is well-defined, then the same must be demanded for $N$, owing to its scalar character. In this case, then \eqref{eq:UH_cond_lapse} implies that $N$ must change sign at the UH. This has an obvious consequence for a black hole -- its interior and exterior are causally reversed. While the region $r>r_{\text {\sc uh}}$ is \emph{future oriented}, the interior $r<r_{\text {\sc uh}}$ is \emph{past oriented}.

Could this be solved in some manner? We can invoke the global time reversal invariance of the solution, meaning that a simultaneous transformation $t\rightarrow -t$ and $U^\m\rightarrow - U^\m$ is also a solution with the same boundary conditions. For this one, the lapse is reversed with respect to that of \eqref{eq:metric} and \eqref{eq:aether}. We can try to glue this solution in the interior of the black hole, so that $N>0$ everywhere. Alas, by doing so, the product $(a\cdot \chi)$ becomes discontinuous, as $N$ approaches zero with non-vanishing derivative from both sides. Hence, continuity and differentiability of $N$ necessarily imply a reversal of the foliation in the interior of the black hole.

At this point one might, and should, be worried about the physical implications of this, since a loss of causality typically implies the existence of closed time loops. We note however that this is not the case here, since the causally reversed region is shielded by the UH. The proper time $t_A$ of any in-falling observer always grows $dt_A>0$, crossing the UH and hitting the singularity. If a second observer could see the in-falling trajectory at all times though, the relation between its time $t_B$ and $t_A$ would flip sign at the UH, having $dt_A/dt_B>0$ for $r>r_{\rm UH}$, but $dt_A/dt_B<0$ for $r<r_{\rm UH}$. However, such an observer does not exist, thus forbidding the existence of time machines and closed time loops. 

\section{Radiation from Universal Horizons}
In the following, we show how our findings are responsible for the emission of Hawking radiation. We consider a Lifshitz scalar field $\phi$ with action in the preferred frame adapted to its motion\footnote{Notice that these rays do not follow geodesics of the metric, but they are however the natural trajectories with no external forces.} \cite{Fradkin:2013sab} 
\begin{align}
    S=-\frac{1}{2}\int d \hat t  dr  \sqrt{|\hat g|}\left(D_{\hat t} \phi D_{\hat t}\phi +\sum_{z=1}^{n}\frac{\alpha_z}{\Lambda^{2z-2}} \phi (-\hat\Delta)^z\phi\right),
\end{align}
where we have suppressed the angular directions, and
\begin{align}
        D_{\hat t}=\frac{1}{\hat N}(\partial_{\hat t} -\hat{N}^i\partial_i),\qquad \hat \Delta=\frac{1}{\hat N}\nabla_i(\hat N \gamma^{ij} \nabla_j).
\end{align}
The hatted quantities are related to those in the ADM metric \eqref{eq:ADM} by the appropriate FDiff transformation \eqref{eq:FDiff} which synchronizes the foliation clock with that of the field.

Close to the UH, both $\gamma^{rr}$ and $\hat N^i$ vanish polynomially in $(r-r_{\text {\sc uh}})$, while $1/\hat N$ diverges. Thus, the equation of motion in that region is 
\begin{align}
\partial_{\hat t}^2\phi+{\cal O}(r-r_{\text {\sc uh}}) \partial_r^2 \phi=0,
\end{align}
whose solutions are $\phi=e^{\pm i \omega \hat{t} +{\cal O}(r-r_{\text {\sc uh}})}$.    

For an observer in the preferred frame, these modes have positive energy whenever we pick the minus sign, while the positive sign leads to negative energy. A sensible quantization and definition of a vacuum $|0\rangle$ can then be done in a standard way. An observer sitting on the preferred frame thus sees a stable vacuum at all times. However, as we will see in a moment, this is not the case for any other observer which does not follow the preferred coordinates\footnote{This explains the results of \cite{Michel:2015rsa}, where the vacuum state is implicitly taken as $|0\rangle$ at all times.}. In particular, we consider here an observer which travels with rays of $\phi$. 

Let us now follow a standard derivation of Hawking radiation \cite{Jacobson:2003vx,Crispino:2007eb}. Consider a positive energy wavepacket of $\phi$, denoted as $\phi_0$, sitting on a leaf approaching the exterior of the UH. 
The lapse $\hat N$ is related to $N$ by
\begin{align}\label{eq:lapse_rel}
    \hat N=\frac{d\tau}{d\hat t}\times  N,
\end{align}
where $d\tau/d \hat t$ is continuous across the UH. Close to $r_{\text {\sc uh}}$ we thus write $d\tau/d\hat t\sim v\equiv \text{constant}$, and
\begin{align}\label{eq:equation_lapse}
    U_\m dx^\m=-\frac{B(r_{\text {\sc uh}}) dr}{\sqrt{-F(r_{\text {\sc uh}})}}=\hat{N} d\hat t =v N d\hat t,
\end{align}
where we have used $N\propto 1+A(r_{\text {\sc uh}})^2F(r_{\text {\sc uh}})=0$.

In order to fix the value of $v$ we follow \cite{Herrero-Valea:2020fqa}. Close to the UH the ray infinitely blue-shifts, its momentum becoming of order $\Lambda$ \cite{Cropp:2013sea}. We thus retain only the highest power in \eqref{eq:disp_relation} $\omega^2\sim \a_{2n} q^{2n}/\Lambda^{2n-2}$. Note however that neither $\omega$ nor $q$ are constant along the trajectory, since they are not constants of motion. Instead, the Killing energy $\Omega=(\chi\cdot k)$ is constant, where $k_\m$ is the four-momentum of the ray. In the preferred frame, it can be decomposed as $k_\m=\omega U_\m -q S_\m$, where $S^\m$ is the space-like unit vector orthogonal to $U^\m$, so we get
\begin{align}
  \Omega=  (\chi\cdot k)=\omega(U\cdot \chi)-q (S\cdot \chi),
\end{align}
with $(S\cdot\chi)=(1-A(r)^2F(r))/(2A(r))$. This provides a second equation that, together with \eqref{eq:disp_relation}, allows to solve for $q(r)$. Working at large $q\sim \Lambda$ and retaining only the leading order, we obtain the group velocity of the ray
\begin{align}
    c(r)=\frac{d\omega}{dq}=\frac{n(1-A(r)^2F(r))}{1+A(r)^2F(r)},
\end{align}
and with it we write its four-velocity $V^\m=U^\m- c(r)S^\m$ \cite{Cropp:2013sea}. Taking the ratio $V^0/V^1$ we thus get
\begin{align}
    \frac{d\hat t}{dr} =\frac{n }{1-n}\frac{B(r_{\text {\sc uh}})}{\sqrt{-F(r_{\text {\sc uh}})}}\frac{1}{N}+{\cal O}(r-r_{\text {\sc uh}}).
\end{align}
Comparing with \eqref{eq:equation_lapse} we read $v=(n-1)/n$.

The preferred time $\hat t$ diverges when approaching $r_{\text {\sc uh}}$. It is related to the radial coordinate in \eqref{eq:ADM} by \eqref{eq:equation_lapse}, so we integrate it when $r\rightarrow r^+_{\text {\sc uh}}$ getting
\begin{align}\label{eq:relation_t_r}
    \hat t=-\frac{\zeta}{\eta}\log[v\eta(r-r_{\text {\sc uh}})]=-\frac{\zeta}{\eta}\log (x)
\end{align}
where we have approximated $N\sim \eta (r-r_{\text {\sc uh}})$ and we have defined $x=v\eta (r-r_{\text {\sc uh}})$ and $\zeta=B(r_{\text {\sc uh}})/(v \sqrt{-F(r_{\text {\sc uh}})})>0$. Notice that since the wavepacket is escaping the UH, $\zeta>0$. The positive mode in the preferred frame will be seen by a free-falling ray crossing the UH as
\begin{align}\label{eq:phi_o}
    \phi_0(x)=e^{-i\omega \hat t}=e^{i\frac{\omega\zeta}{\eta}\log(x) },
\end{align}
and will in general contain positive and negative modes with respect to the vacuum defined by this observer $|0\rangle_{\rm ff}$, which sees frequencies $\bar\omega$. In order to disentangle both contributions, we follow Unruh \cite{Unruh:1976db}, by noting that any function bounded as $|x| \rightarrow \infty$ and analytic in the lower half of the complex plane contains only positive modes\footnote{This is a consequence of Fourier decomposition, since the Fourier measure $e^{-i\bar \omega x}$ diverges at large $x$ for negative $\bar\omega$, unless $\phi_0(\bar\omega)$ vanishes for $\bar\omega<0$.} \cite{Jacobson:2003vx}.

To extract the positive part of $\phi_0$ we thus perform an analytic continuation of $\log(x)$ to $x<0$. Since we want the function to be analytic in the lower half of the complex plane, we put the branch cut in the upper half, and fix the continuation to $\log(-x)+i\pi$. With this, we have a function which is fully analytic in the lower half. The positive frequency part of the mode must be proportional to it
\begin{align}
    \phi^+=c_+\left( \phi_0(x) + \phi_0(-x)e^{-\frac{\pi \zeta \omega}{\eta}}\right),
\end{align}
with $c_+$ constant. The negative frequency part is given by choosing an extension analytic in the upper part of the complex plane, by $\log(-x)-i\pi$ and thus
\begin{align}
\phi^-=c_- \left( \phi_0(x) +\phi_0(-x)e^{\frac{\pi \zeta \omega}{\eta}}\right).
\end{align}
The constants $c_\pm$ are fixed by demanding $\phi^++\phi^-=\phi_0$ and read $c_-=\left(1-e^{\frac{2\pi \omega\zeta}{\eta}}\right)^{-1}$ and $c_+=-c_- e^{-\frac{2\pi \omega\zeta}{\eta}}$.

The average number of particles seen by an observer in free fall with the field ray is given by $\langle {\cal N}\rangle=-\langle \phi^-\phi^-\rangle$, where ${\cal N}$ is the particle number operator. In order to compute it, we define the internal product of states in a foliation leaf
\begin{align}
    \langle \psi_1\psi_2\rangle_{\hat t} =i\int \sqrt{|\gamma|}\ dr\ \left(\psi^\dagger_1 \partial_{\hat t} \psi_2-(\partial_{\hat t} \psi^\dagger_1)\psi_2\right),
\end{align}
and note that $\langle\phi_0(x)\phi_0(-x)\rangle=0$, since the functions have disjoint support. We also note that $\phi_0(-x)$ corresponds to a mode in the region $r<r_{\rm UH}$. Thus, in a similar form to \eqref{eq:relation_t_r} and \eqref{eq:phi_o} we rewrite the positive mode $e^{-i \omega \hat t}$ in terms of $x<0$, by integrating the lapse close to $r_{\rm UH}^-$. In order to proof our point about the need for reversing the interior foliation, we expand $N$ now as $N\sim \pm \eta(r-r_{\text {\sc uh}})$. The positive sign corresponds to a reversed interior, while the minus sign refers to the gluing of two solutions which keeps $N>0$ everywhere.

Comparing both cases we get
\begin{align}
    \left.\frac{d\hat t}{dr}\right|_{x>0}=\mp \left.\frac{d\hat t}{dr}\right|_{x<0}.
\end{align}

Taking this into account, we note that $\langle \phi_0(x)\phi_0(x)\rangle=\mp \langle \phi_0(-x)\phi_0(-x)\rangle$ by direct comparison of the products. We finally arrive to 
\begin{align}\label{eq:Np}
    \langle {\cal N}\rangle =-\langle \phi_0(x)\phi_0(x)\rangle\left(1\mp e^{\frac{2\pi \omega\zeta}{\eta}}\right)\left(1-e^{\frac{2\pi \omega\zeta}{\eta}}\right)^{-2}.
\end{align}

As we argued, choosing the sign that \emph{keeps} the causal ordering in both regions identical, which corresponds to the positive sign in $\langle {\cal N}\rangle$, leads to fundamental problems. The average number of particles \eqref{eq:Np} is negative for all $\omega>0$, which implies a loss of unitarity.\footnote{Let us stress that such a loss of unitarity is much more dramatic than the one usually associated to the information loss problem in black hole physics. While the latter comes from assuming complete thermal evaporation of the black hole, which in turn requires tracing over the Hawking partners in its interior (so turning a pure state into a mixed one), the second is truly associated to the violation of a basic tenet of quantum mechanics.} This is easily seen through the density matrix operator $\rho$, because
\begin{align}
    \langle {\cal N}\rangle=\frac{1}{Z}\text{Tr}\left(\rho \cdot {\cal N}\right).
\end{align}
Since both ${\cal N}$ and the partition function $Z$ are semi-positive definite, $\langle {\cal N}\rangle<0$ implies that there exist some eigenstates $n_i$ of ${\cal N}$ for which $\langle n_i|\rho|n_i\rangle<0$, violating the axioms of probability. 

We conclude that a reversal of the foliation in the interior of the black hole is a necessary condition for the quantum field theory \emph{in the exterior} to be well defined. By choosing the correct sign we thus arrive to 
\begin{align}
     \langle {\cal N}\rangle =\langle \phi_0(x)\phi_0(x)\rangle\left(e^{\frac{2\pi \omega\zeta}{\eta}}-1\right)^{-1}.
\end{align}
This is a thermal distribution of the Bose-Einstein type, with greybody factor $\langle \phi_0(x)\phi_0(x)\rangle$ and temperature
\begin{align}\label{eq:temperature}
& T=\frac{\eta}{2\pi \zeta}=\frac{n-1}{n}\frac{\kappa_{\text {\sc uh}}}{\pi},
\end{align}
where we have computed $\eta=\left.\partial_r N\right|_{r_{\text {\sc uh}}}$ from \eqref{eq:adotchi} and $2\kappa_{\text {\sc uh}}=-\left.(a\cdot\chi)\right|_{\text {\sc uh}}$. As we observe, we recover expression \eqref{eq:Hawking_temp}.

\section{Temperature of UHs}
Let us now consider explicit examples of analytic solutions to the action of EA gravity \cite{Jacobson:2000xp}
\begin{align}
    S=-\frac{1}{16\pi G}\int d^4x \sqrt{|g|} \left(R+{\cal L}_U + \lambda(U_\m U^\m-1)\right),
\end{align}
where $\lambda$ is a Lagrange multiplier and ${\cal L}_U=K^{\a\b}_{\m\n}\nabla_\a U^\m \nabla_\b U^\n$, with $K^{\a\b}_{\m\n}=c_1 g^{\a\b}g_{\m\n}+c_2 \delta^\a_\m \delta^\b_\n +c_3 \delta^\a_\n \delta^\n_\m + c_4 U^\a U^\b g_{\m\n}$ and couplings $c_i$. 

The solutions that we consider were obtained in \cite{Berglund:2012bu} and for both $B(r)=1$. However, they only solve the equations of motion for particular values of $c_i$. They are\\
\phantom{aa}\\
\emph{Solution 1:} $c_1+c_2+c_3=0$,
    \begin{align}\label{eq:sol_1}
       &A(r)=\frac{1}{1+\frac{{\cal R}}{r}},  \ F(r)=1-\frac{2GM}{r}-\frac{{\cal R}(2GM+{\cal R})}{r^2},
          \end{align}
     where ${\cal R}=GM\left(\sqrt{\frac{2-c_1-c_4}{2(1-c_1-c_3)}}-1\right)$ and $r_{\text {\sc uh}}=GM$.\\
\phantom{aa}\\
\emph{Solution 2:} $c_1+c_4=0$
    \begin{align}\label{eq:sol_2}
    \nonumber     &F(r)=1-\frac{2GM}{r} - \frac{(c_1+c_3)r_u^4}{r^4},\\
         &A(r)=\frac{1}{F(r)}\left(-\frac{r_{u}^2}{r^2}+\sqrt{F(r)+\frac{r_u^4}{r^4}}\right),
    \end{align}
where now $r_u=\frac{GM}{2}\left(\frac{27}{1-c_1-c_3}\right)^{\frac{1}{4}}$ and $r_{\text {\sc uh}}=\frac{3GM}{2}$. In both cases $M$ is the Komar mass of the black hole.

Plugging these solutions into \eqref{eq:temperature} we find
\begin{align}
    T_{1}=\frac{n-1}{n}\frac{1}{2\sqrt{2}\ \pi G M }\sqrt{\frac{2-c_1-c_4}{1-c_1-c_3}},\\
    T_{2}=\frac{n-1}{n}\frac{1}{3\sqrt{3}\ \pi GM }\sqrt{\frac{2}{1-c_1-c_3}}.
\end{align}

A particular limit of \eqref{eq:sol_1} -- where $c_1+c_4=c_1+c_3=0$ -- was explored in \cite{Herrero-Valea:2020fqa}. Comparing with them, we find perfect agreement for the same values of the couplings. However, our result here is valid for any $c_i$, and it is not restricted to a Schwarzschild metric. Nevertheless, this corner of the parameter space has been already ruled out by observations \cite{Gupta:2021vdj}. On the other hand, \eqref{eq:sol_2} can be made compatible with observational bounds by simply letting $c_1+c_3=0$, which fixes the speed of gravitational waves to $c=1$. In this case, \eqref{eq:sol_2} is the only non-rotating spherically symmetric solution \emph{with a regular UH}, and it seems to be the asymptotic endpoint of a spherical collapse with asymptotically flat boundary conditions \cite{Franchini:2021bpt}.

\section{Conclusions}
In this \emph{paper} we have shown that continuity and differentiability of $N$ through a UH requires the interior foliation of a black hole to be reversed with respect to the exterior one. Although this is harmless in terms of allowing an observer to travel along a closed time loop, it has strong consequences for the quantum properties of gravitating fields. There is a strong dependence on the analytic continuation of the positive frequency modes across the UH. 

In well-defined quantum theories, this continuation is chosen such that the mode is single-valued through the UH. However, lapses with a kink there fail to satisfy this property. This seems to result in thermodynamic properties. We have shown how UHs radiate particles with Hawking temperature \eqref{eq:temperature} only if the reversed interior is allowed. Our results apply to all stationary and spherically symmetric spacetimes. If the lapse is forced to be positive everywhere, violations of unitarity are found instead. There seems to be a strict relation between the well-posedness of the QFT and thermal properties of BHs. Future investigations will show if such connection endures detailed phenomenological and conceptual analyses which we have already started in this paper. 

To this aim, we have applied our results to known analytic solutions. Besides purely theoretically interesting spacetimes, these also include \eqref{eq:sol_2}, which is compatible with all current observations when $c_1+c_3=0$. The need for a reversal of the interior foliation sheds a shadow on the actual physical validity of this solution in astrophysical settings, however. It seems hard to develop such a strongly non-local property from local evolution of collapsing matter. Absence of a Birkhoff theorem in Lorentz violating gravity leaves the door open for other, perhaps similar, solutions to be realized in Nature.

Our findings here also highlight the significance of the dynamics of the foliation orthogonal to $U^\m$, whose importance is sometimes overlooked when the four-dimensional diffeomorphism invariant description is used. It would be interesting to understand if shifting the perspective could allow for finding rotating black hole solutions with a UH, which have been elusive so far\cite{Adam:2021vsk}.

A final question for the future regards the stability of the UH. Since the modes are infinitely blue-shifted when approaching $r_{\rm UH}$, one could contest the truncation of the action to second derivatives. Higher derivative terms are required in Ho\v rava gravity to have a regular UV behavior \cite{Barvinsky:2015kil,Barvinsky:2017kob,Griffin:2017wvh,Barvinsky:2021ubv}, and we could wonder if they have a role in the dynamics of the UH.

\section*{Acknowledgements}
We are grateful to Enrico Barausse, David Mattingly and Raquel Santos-Garc\' ia for comments and discussions. Our work has been supported by the Italian Ministry of
Education and Scientific Research (MIUR) under the grant PRIN MIUR 2017-MB8AEZ.
\bibliography{biblio}{}

\begin{thebibliography}{30}%
\makeatletter
\providecommand \@ifxundefined [1]{%
 \@ifx{#1\undefined}
}%
\providecommand \@ifnum [1]{%
 \ifnum #1\expandafter \@firstoftwo
 \else \expandafter \@secondoftwo
 \fi
}%
\providecommand \@ifx [1]{%
 \ifx #1\expandafter \@firstoftwo
 \else \expandafter \@secondoftwo
 \fi
}%
\providecommand \natexlab [1]{#1}%
\providecommand \enquote  [1]{``#1''}%
\providecommand \bibnamefont  [1]{#1}%
\providecommand \bibfnamefont [1]{#1}%
\providecommand \citenamefont [1]{#1}%
\providecommand \href@noop [0]{\@secondoftwo}%
\providecommand \href [0]{\begingroup \@sanitize@url \@href}%
\providecommand \@href[1]{\@@startlink{#1}\@@href}%
\providecommand \@@href[1]{\endgroup#1\@@endlink}%
\providecommand \@sanitize@url [0]{\catcode `\\12\catcode `\$12\catcode
  `\&12\catcode `\#12\catcode `\^12\catcode `\_12\catcode `\%12\relax}%
\providecommand \@@startlink[1]{}%
\providecommand \@@endlink[0]{}%
\providecommand \url  [0]{\begingroup\@sanitize@url \@url }%
\providecommand \@url [1]{\endgroup\@href {#1}{\urlprefix }}%
\providecommand \urlprefix  [0]{URL }%
\providecommand \Eprint [0]{\href }%
\providecommand \doibase [0]{https://doi.org/}%
\providecommand \selectlanguage [0]{\@gobble}%
\providecommand \bibinfo  [0]{\@secondoftwo}%
\providecommand \bibfield  [0]{\@secondoftwo}%
\providecommand \translation [1]{[#1]}%
\providecommand \BibitemOpen [0]{}%
\providecommand \bibitemStop [0]{}%
\providecommand \bibitemNoStop [0]{.\EOS\space}%
\providecommand \EOS [0]{\spacefactor3000\relax}%
\providecommand \BibitemShut  [1]{\csname bibitem#1\endcsname}%
\let\auto@bib@innerbib\@empty
\bibitem [{\citenamefont {Jacobson}(2005)}]{Jacobson:2003vx}%
  \BibitemOpen
  \bibfield  {author} {\bibinfo {author} {\bibfnamefont {T.}~\bibnamefont
  {Jacobson}},\ }\bibinfo {title} {Introduction to quantum fields in curved
  spacetime and the hawking effect},\ in\ \href
  {https://doi.org/10.1007/0-387-24992-3_2} {\emph {\bibinfo {booktitle}
  {Lectures on Quantum Gravity}}},\ \bibinfo {editor} {edited by\ \bibinfo
  {editor} {\bibfnamefont {A.}~\bibnamefont {Gomberoff}}\ and\ \bibinfo
  {editor} {\bibfnamefont {D.}~\bibnamefont {Marolf}}}\ (\bibinfo  {publisher}
  {Springer US},\ \bibinfo {address} {Boston, MA},\ \bibinfo {year} {2005})\
  pp.\ \bibinfo {pages} {39--89}\BibitemShut {NoStop}%
\bibitem [{\citenamefont {Jacobson}\ and\ \citenamefont
  {Mattingly}(2001)}]{Jacobson:2000xp}%
  \BibitemOpen
  \bibfield  {author} {\bibinfo {author} {\bibfnamefont {T.}~\bibnamefont
  {Jacobson}}\ and\ \bibinfo {author} {\bibfnamefont {D.}~\bibnamefont
  {Mattingly}},\ }\bibfield  {title} {\bibinfo {title} {{Gravity with a
  dynamical preferred frame}},\ }\href
  {https://doi.org/10.1103/PhysRevD.64.024028} {\bibfield  {journal} {\bibinfo
  {journal} {Phys. Rev. D}\ }\textbf {\bibinfo {volume} {64}},\ \bibinfo
  {pages} {024028} (\bibinfo {year} {2001})},\ \Eprint
  {https://arxiv.org/abs/gr-qc/0007031} {arXiv:gr-qc/0007031} \BibitemShut
  {NoStop}%
\bibitem [{\citenamefont {Berglund}\ \emph {et~al.}(2012)\citenamefont
  {Berglund}, \citenamefont {Bhattacharyya},\ and\ \citenamefont
  {Mattingly}}]{Berglund:2012bu}%
  \BibitemOpen
  \bibfield  {author} {\bibinfo {author} {\bibfnamefont {P.}~\bibnamefont
  {Berglund}}, \bibinfo {author} {\bibfnamefont {J.}~\bibnamefont
  {Bhattacharyya}},\ and\ \bibinfo {author} {\bibfnamefont {D.}~\bibnamefont
  {Mattingly}},\ }\bibfield  {title} {\bibinfo {title} {{Mechanics of universal
  horizons}},\ }\href {https://doi.org/10.1103/PhysRevD.85.124019} {\bibfield
  {journal} {\bibinfo  {journal} {Phys. Rev. D}\ }\textbf {\bibinfo {volume}
  {85}},\ \bibinfo {pages} {124019} (\bibinfo {year} {2012})},\ \Eprint
  {https://arxiv.org/abs/1202.4497} {arXiv:1202.4497 [hep-th]} \BibitemShut
  {NoStop}%
\bibitem [{\citenamefont {Barausse}\ and\ \citenamefont
  {Sotiriou}(2013)}]{Barausse:2012qh}%
  \BibitemOpen
  \bibfield  {author} {\bibinfo {author} {\bibfnamefont {E.}~\bibnamefont
  {Barausse}}\ and\ \bibinfo {author} {\bibfnamefont {T.~P.}\ \bibnamefont
  {Sotiriou}},\ }\bibfield  {title} {\bibinfo {title} {{Slowly rotating black
  holes in Horava-Lifshitz gravity}},\ }\href
  {https://doi.org/10.1103/PhysRevD.87.087504} {\bibfield  {journal} {\bibinfo
  {journal} {Phys. Rev. D}\ }\textbf {\bibinfo {volume} {87}},\ \bibinfo
  {pages} {087504} (\bibinfo {year} {2013})},\ \Eprint
  {https://arxiv.org/abs/1212.1334} {arXiv:1212.1334 [gr-qc]} \BibitemShut
  {NoStop}%
\bibitem [{\citenamefont {Barausse}\ \emph {et~al.}(2016)\citenamefont
  {Barausse}, \citenamefont {Sotiriou},\ and\ \citenamefont
  {Vega}}]{Barausse:2015frm}%
  \BibitemOpen
  \bibfield  {author} {\bibinfo {author} {\bibfnamefont {E.}~\bibnamefont
  {Barausse}}, \bibinfo {author} {\bibfnamefont {T.~P.}\ \bibnamefont
  {Sotiriou}},\ and\ \bibinfo {author} {\bibfnamefont {I.}~\bibnamefont
  {Vega}},\ }\bibfield  {title} {\bibinfo {title} {{Slowly rotating black holes
  in Einstein-\ae{}ther theory}},\ }\href
  {https://doi.org/10.1103/PhysRevD.93.044044} {\bibfield  {journal} {\bibinfo
  {journal} {Phys. Rev. D}\ }\textbf {\bibinfo {volume} {93}},\ \bibinfo
  {pages} {044044} (\bibinfo {year} {2016})},\ \Eprint
  {https://arxiv.org/abs/1512.05894} {arXiv:1512.05894 [gr-qc]} \BibitemShut
  {NoStop}%
\bibitem [{\citenamefont {Horava}(2009)}]{Horava:2009uw}%
  \BibitemOpen
  \bibfield  {author} {\bibinfo {author} {\bibfnamefont {P.}~\bibnamefont
  {Horava}},\ }\bibfield  {title} {\bibinfo {title} {{Quantum Gravity at a
  Lifshitz Point}},\ }\href {https://doi.org/10.1103/PhysRevD.79.084008}
  {\bibfield  {journal} {\bibinfo  {journal} {Phys. Rev. D}\ }\textbf {\bibinfo
  {volume} {79}},\ \bibinfo {pages} {084008} (\bibinfo {year} {2009})},\
  \Eprint {https://arxiv.org/abs/0901.3775} {arXiv:0901.3775 [hep-th]}
  \BibitemShut {NoStop}%
\bibitem [{\citenamefont {Blas}\ \emph {et~al.}(2011)\citenamefont {Blas},
  \citenamefont {Pujolas},\ and\ \citenamefont {Sibiryakov}}]{Blas:2010hb}%
  \BibitemOpen
  \bibfield  {author} {\bibinfo {author} {\bibfnamefont {D.}~\bibnamefont
  {Blas}}, \bibinfo {author} {\bibfnamefont {O.}~\bibnamefont {Pujolas}},\ and\
  \bibinfo {author} {\bibfnamefont {S.}~\bibnamefont {Sibiryakov}},\ }\bibfield
   {title} {\bibinfo {title} {{Models of non-relativistic quantum gravity: The
  Good, the bad and the healthy}},\ }\href
  {https://doi.org/10.1007/JHEP04(2011)018} {\bibfield  {journal} {\bibinfo
  {journal} {JHEP}\ }\textbf {\bibinfo {volume} {04}},\ \bibinfo {pages}
  {018}},\ \Eprint {https://arxiv.org/abs/1007.3503} {arXiv:1007.3503 [hep-th]}
  \BibitemShut {NoStop}%
\bibitem [{\citenamefont {Blas}\ and\ \citenamefont
  {Sibiryakov}(2011)}]{Blas:2011ni}%
  \BibitemOpen
  \bibfield  {author} {\bibinfo {author} {\bibfnamefont {D.}~\bibnamefont
  {Blas}}\ and\ \bibinfo {author} {\bibfnamefont {S.}~\bibnamefont
  {Sibiryakov}},\ }\bibfield  {title} {\bibinfo {title} {{Horava gravity versus
  thermodynamics: The Black hole case}},\ }\href
  {https://doi.org/10.1103/PhysRevD.84.124043} {\bibfield  {journal} {\bibinfo
  {journal} {Phys. Rev. D}\ }\textbf {\bibinfo {volume} {84}},\ \bibinfo
  {pages} {124043} (\bibinfo {year} {2011})},\ \Eprint
  {https://arxiv.org/abs/1110.2195} {arXiv:1110.2195 [hep-th]} \BibitemShut
  {NoStop}%
\bibitem [{\citenamefont {Berglund}\ \emph {et~al.}(2013)\citenamefont
  {Berglund}, \citenamefont {Bhattacharyya},\ and\ \citenamefont
  {Mattingly}}]{Berglund:2012fk}%
  \BibitemOpen
  \bibfield  {author} {\bibinfo {author} {\bibfnamefont {P.}~\bibnamefont
  {Berglund}}, \bibinfo {author} {\bibfnamefont {J.}~\bibnamefont
  {Bhattacharyya}},\ and\ \bibinfo {author} {\bibfnamefont {D.}~\bibnamefont
  {Mattingly}},\ }\bibfield  {title} {\bibinfo {title} {{Towards Thermodynamics
  of Universal Horizons in Einstein-\ae{}ther Theory}},\ }\href
  {https://doi.org/10.1103/PhysRevLett.110.071301} {\bibfield  {journal}
  {\bibinfo  {journal} {Phys. Rev. Lett.}\ }\textbf {\bibinfo {volume} {110}},\
  \bibinfo {pages} {071301} (\bibinfo {year} {2013})},\ \Eprint
  {https://arxiv.org/abs/1210.4940} {arXiv:1210.4940 [hep-th]} \BibitemShut
  {NoStop}%
\bibitem [{\citenamefont {Pacilio}\ and\ \citenamefont
  {Liberati}(2017)}]{Pacilio:2017swi}%
  \BibitemOpen
  \bibfield  {author} {\bibinfo {author} {\bibfnamefont {C.}~\bibnamefont
  {Pacilio}}\ and\ \bibinfo {author} {\bibfnamefont {S.}~\bibnamefont
  {Liberati}},\ }\bibfield  {title} {\bibinfo {title} {{First law of black
  holes with a universal horizon}},\ }\href
  {https://doi.org/10.1103/PhysRevD.96.104060} {\bibfield  {journal} {\bibinfo
  {journal} {Phys. Rev. D}\ }\textbf {\bibinfo {volume} {96}},\ \bibinfo
  {pages} {104060} (\bibinfo {year} {2017})},\ \Eprint
  {https://arxiv.org/abs/1709.05802} {arXiv:1709.05802 [gr-qc]} \BibitemShut
  {NoStop}%
\bibitem [{\citenamefont {Ding}\ \emph {et~al.}(2016)\citenamefont {Ding},
  \citenamefont {Wang}, \citenamefont {Wang},\ and\ \citenamefont
  {Zhu}}]{Ding:2015fyx}%
  \BibitemOpen
  \bibfield  {author} {\bibinfo {author} {\bibfnamefont {C.}~\bibnamefont
  {Ding}}, \bibinfo {author} {\bibfnamefont {A.}~\bibnamefont {Wang}}, \bibinfo
  {author} {\bibfnamefont {X.}~\bibnamefont {Wang}},\ and\ \bibinfo {author}
  {\bibfnamefont {T.}~\bibnamefont {Zhu}},\ }\bibfield  {title} {\bibinfo
  {title} {{Hawking radiation of charged Einstein-aether black holes at both
  Killing and universal horizons}},\ }\href
  {https://doi.org/10.1016/j.nuclphysb.2016.10.007} {\bibfield  {journal}
  {\bibinfo  {journal} {Nucl. Phys. B}\ }\textbf {\bibinfo {volume} {913}},\
  \bibinfo {pages} {694} (\bibinfo {year} {2016})},\ \Eprint
  {https://arxiv.org/abs/1512.01900} {arXiv:1512.01900 [gr-qc]} \BibitemShut
  {NoStop}%
\bibitem [{\citenamefont {Ding}\ and\ \citenamefont
  {Liu}(2017)}]{Ding:2016srk}%
  \BibitemOpen
  \bibfield  {author} {\bibinfo {author} {\bibfnamefont {C.}~\bibnamefont
  {Ding}}\ and\ \bibinfo {author} {\bibfnamefont {C.}~\bibnamefont {Liu}},\
  }\bibfield  {title} {\bibinfo {title} {{Dispersion relation and surface
  gravity of universal horizons}},\ }\href
  {https://doi.org/10.1007/s11433-017-9012-8} {\bibfield  {journal} {\bibinfo
  {journal} {Sci. China Phys. Mech. Astron.}\ }\textbf {\bibinfo {volume}
  {60}},\ \bibinfo {pages} {050411} (\bibinfo {year} {2017})},\ \Eprint
  {https://arxiv.org/abs/1611.03153} {arXiv:1611.03153 [gr-qc]} \BibitemShut
  {NoStop}%
\bibitem [{\citenamefont {Cropp}(2016)}]{Cropp:2016gkn}%
  \BibitemOpen
  \bibfield  {author} {\bibinfo {author} {\bibfnamefont {B.}~\bibnamefont
  {Cropp}},\ }\emph {\bibinfo {title} {{Strange Horizons: Understanding Causal
  Barriers Beyond General Relativity}}},\ \href@noop {} {\bibinfo {type} {Phd
  thesis}},\ \bibinfo  {school} {SISSA} (\bibinfo {year} {2016}),\ \Eprint
  {https://arxiv.org/abs/1611.00208} {arXiv:1611.00208 [gr-qc]} \BibitemShut
  {NoStop}%
\bibitem [{\citenamefont {Herrero-Valea}\ \emph {et~al.}(2021)\citenamefont
  {Herrero-Valea}, \citenamefont {Liberati},\ and\ \citenamefont
  {Santos-Garcia}}]{Herrero-Valea:2020fqa}%
  \BibitemOpen
  \bibfield  {author} {\bibinfo {author} {\bibfnamefont {M.}~\bibnamefont
  {Herrero-Valea}}, \bibinfo {author} {\bibfnamefont {S.}~\bibnamefont
  {Liberati}},\ and\ \bibinfo {author} {\bibfnamefont {R.}~\bibnamefont
  {Santos-Garcia}},\ }\bibfield  {title} {\bibinfo {title} {{Hawking Radiation
  from Universal Horizons}},\ }\href {https://doi.org/10.1007/JHEP04(2021)255}
  {\bibfield  {journal} {\bibinfo  {journal} {JHEP}\ }\textbf {\bibinfo
  {volume} {04}},\ \bibinfo {pages} {255}},\ \Eprint
  {https://arxiv.org/abs/2101.00028} {arXiv:2101.00028 [gr-qc]} \BibitemShut
  {NoStop}%
\bibitem [{\citenamefont {Michel}\ and\ \citenamefont
  {Parentani}(2015)}]{Michel:2015rsa}%
  \BibitemOpen
  \bibfield  {author} {\bibinfo {author} {\bibfnamefont {F.}~\bibnamefont
  {Michel}}\ and\ \bibinfo {author} {\bibfnamefont {R.}~\bibnamefont
  {Parentani}},\ }\bibfield  {title} {\bibinfo {title} {{Black hole radiation
  in the presence of a universal horizon}},\ }\href
  {https://doi.org/10.1103/PhysRevD.91.124049} {\bibfield  {journal} {\bibinfo
  {journal} {Phys. Rev. D}\ }\textbf {\bibinfo {volume} {91}},\ \bibinfo
  {pages} {124049} (\bibinfo {year} {2015})},\ \Eprint
  {https://arxiv.org/abs/1505.00332} {arXiv:1505.00332 [gr-qc]} \BibitemShut
  {NoStop}%
\bibitem [{\citenamefont {Pospelov}\ and\ \citenamefont
  {Shang}(2012)}]{Pospelov:2010mp}%
  \BibitemOpen
  \bibfield  {author} {\bibinfo {author} {\bibfnamefont {M.}~\bibnamefont
  {Pospelov}}\ and\ \bibinfo {author} {\bibfnamefont {Y.}~\bibnamefont
  {Shang}},\ }\bibfield  {title} {\bibinfo {title} {{On Lorentz violation in
  Horava-Lifshitz type theories}},\ }\href
  {https://doi.org/10.1103/PhysRevD.85.105001} {\bibfield  {journal} {\bibinfo
  {journal} {Phys. Rev. D}\ }\textbf {\bibinfo {volume} {85}},\ \bibinfo
  {pages} {105001} (\bibinfo {year} {2012})},\ \Eprint
  {https://arxiv.org/abs/1010.5249} {arXiv:1010.5249 [hep-th]} \BibitemShut
  {NoStop}%
\bibitem [{\citenamefont {Bhattacharyya}\ \emph {et~al.}(2016)\citenamefont
  {Bhattacharyya}, \citenamefont {Colombo},\ and\ \citenamefont
  {Sotiriou}}]{Bhattacharyya:2015gwa}%
  \BibitemOpen
  \bibfield  {author} {\bibinfo {author} {\bibfnamefont {J.}~\bibnamefont
  {Bhattacharyya}}, \bibinfo {author} {\bibfnamefont {M.}~\bibnamefont
  {Colombo}},\ and\ \bibinfo {author} {\bibfnamefont {T.~P.}\ \bibnamefont
  {Sotiriou}},\ }\bibfield  {title} {\bibinfo {title} {{Causality and black
  holes in spacetimes with a preferred foliation}},\ }\href
  {https://doi.org/10.1088/0264-9381/33/23/235003} {\bibfield  {journal}
  {\bibinfo  {journal} {Class. Quant. Grav.}\ }\textbf {\bibinfo {volume}
  {33}},\ \bibinfo {pages} {235003} (\bibinfo {year} {2016})},\ \Eprint
  {https://arxiv.org/abs/1509.01558} {arXiv:1509.01558 [gr-qc]} \BibitemShut
  {NoStop}%
\bibitem [{\citenamefont {Arnowitt}\ \emph {et~al.}(2008)\citenamefont
  {Arnowitt}, \citenamefont {Deser},\ and\ \citenamefont
  {Misner}}]{Arnowitt:1962hi}%
  \BibitemOpen
  \bibfield  {author} {\bibinfo {author} {\bibfnamefont {R.~L.}\ \bibnamefont
  {Arnowitt}}, \bibinfo {author} {\bibfnamefont {S.}~\bibnamefont {Deser}},\
  and\ \bibinfo {author} {\bibfnamefont {C.~W.}\ \bibnamefont {Misner}},\
  }\bibfield  {title} {\bibinfo {title} {{The Dynamics of general
  relativity}},\ }\href {https://doi.org/10.1007/s10714-008-0661-1} {\bibfield
  {journal} {\bibinfo  {journal} {Gen. Rel. Grav.}\ }\textbf {\bibinfo {volume}
  {40}},\ \bibinfo {pages} {1997} (\bibinfo {year} {2008})},\ \Eprint
  {https://arxiv.org/abs/gr-qc/0405109} {arXiv:gr-qc/0405109} \BibitemShut
  {NoStop}%
\bibitem [{\citenamefont {Liberati}(2013)}]{Liberati:2013xla}%
  \BibitemOpen
  \bibfield  {author} {\bibinfo {author} {\bibfnamefont {S.}~\bibnamefont
  {Liberati}},\ }\bibfield  {title} {\bibinfo {title} {{Tests of Lorentz
  invariance: a 2013 update}},\ }\href
  {https://doi.org/10.1088/0264-9381/30/13/133001} {\bibfield  {journal}
  {\bibinfo  {journal} {Class. Quant. Grav.}\ }\textbf {\bibinfo {volume}
  {30}},\ \bibinfo {pages} {133001} (\bibinfo {year} {2013})},\ \Eprint
  {https://arxiv.org/abs/1304.5795} {arXiv:1304.5795 [gr-qc]} \BibitemShut
  {NoStop}%
\bibitem [{\citenamefont {Fradkin}(2013)}]{Fradkin:2013sab}%
  \BibitemOpen
  \bibfield  {author} {\bibinfo {author} {\bibfnamefont {E.~H.}\ \bibnamefont
  {Fradkin}},\ }\href@noop {} {\emph {\bibinfo {title} {{Field Theories of
  Condensed Matter Physics}}}},\ Vol.~\bibinfo {volume} {82}\ (\bibinfo
  {publisher} {Cambridge Univ. Press},\ \bibinfo {address} {Cambridge, UK},\
  \bibinfo {year} {2013})\BibitemShut {NoStop}%
\bibitem [{\citenamefont {Crispino}\ \emph {et~al.}(2008)\citenamefont
  {Crispino}, \citenamefont {Higuchi},\ and\ \citenamefont
  {Matsas}}]{Crispino:2007eb}%
  \BibitemOpen
  \bibfield  {author} {\bibinfo {author} {\bibfnamefont {L.~C.~B.}\
  \bibnamefont {Crispino}}, \bibinfo {author} {\bibfnamefont {A.}~\bibnamefont
  {Higuchi}},\ and\ \bibinfo {author} {\bibfnamefont {G.~E.~A.}\ \bibnamefont
  {Matsas}},\ }\bibfield  {title} {\bibinfo {title} {{The Unruh effect and its
  applications}},\ }\href {https://doi.org/10.1103/RevModPhys.80.787}
  {\bibfield  {journal} {\bibinfo  {journal} {Rev. Mod. Phys.}\ }\textbf
  {\bibinfo {volume} {80}},\ \bibinfo {pages} {787} (\bibinfo {year} {2008})},\
  \Eprint {https://arxiv.org/abs/0710.5373} {arXiv:0710.5373 [gr-qc]}
  \BibitemShut {NoStop}%
\bibitem [{\citenamefont {Cropp}\ \emph {et~al.}(2014)\citenamefont {Cropp},
  \citenamefont {Liberati}, \citenamefont {Mohd},\ and\ \citenamefont
  {Visser}}]{Cropp:2013sea}%
  \BibitemOpen
  \bibfield  {author} {\bibinfo {author} {\bibfnamefont {B.}~\bibnamefont
  {Cropp}}, \bibinfo {author} {\bibfnamefont {S.}~\bibnamefont {Liberati}},
  \bibinfo {author} {\bibfnamefont {A.}~\bibnamefont {Mohd}},\ and\ \bibinfo
  {author} {\bibfnamefont {M.}~\bibnamefont {Visser}},\ }\bibfield  {title}
  {\bibinfo {title} {{Ray tracing Einstein-\AE{}ther black holes: Universal
  versus Killing horizons}},\ }\href
  {https://doi.org/10.1103/PhysRevD.89.064061} {\bibfield  {journal} {\bibinfo
  {journal} {Phys. Rev. D}\ }\textbf {\bibinfo {volume} {89}},\ \bibinfo
  {pages} {064061} (\bibinfo {year} {2014})},\ \Eprint
  {https://arxiv.org/abs/1312.0405} {arXiv:1312.0405 [gr-qc]} \BibitemShut
  {NoStop}%
\bibitem [{\citenamefont {Unruh}(1976)}]{Unruh:1976db}%
  \BibitemOpen
  \bibfield  {author} {\bibinfo {author} {\bibfnamefont {W.~G.}\ \bibnamefont
  {Unruh}},\ }\bibfield  {title} {\bibinfo {title} {{Notes on black hole
  evaporation}},\ }\href {https://doi.org/10.1103/PhysRevD.14.870} {\bibfield
  {journal} {\bibinfo  {journal} {Phys. Rev. D}\ }\textbf {\bibinfo {volume}
  {14}},\ \bibinfo {pages} {870} (\bibinfo {year} {1976})}\BibitemShut
  {NoStop}%
\bibitem [{\citenamefont {Gupta}\ \emph {et~al.}(2021)\citenamefont {Gupta},
  \citenamefont {Herrero-Valea}, \citenamefont {Blas}, \citenamefont
  {Barausse}, \citenamefont {Cornish}, \citenamefont {Yagi},\ and\
  \citenamefont {Yunes}}]{Gupta:2021vdj}%
  \BibitemOpen
  \bibfield  {author} {\bibinfo {author} {\bibfnamefont {T.}~\bibnamefont
  {Gupta}}, \bibinfo {author} {\bibfnamefont {M.}~\bibnamefont
  {Herrero-Valea}}, \bibinfo {author} {\bibfnamefont {D.}~\bibnamefont {Blas}},
  \bibinfo {author} {\bibfnamefont {E.}~\bibnamefont {Barausse}}, \bibinfo
  {author} {\bibfnamefont {N.}~\bibnamefont {Cornish}}, \bibinfo {author}
  {\bibfnamefont {K.}~\bibnamefont {Yagi}},\ and\ \bibinfo {author}
  {\bibfnamefont {N.}~\bibnamefont {Yunes}},\ }\bibfield  {title} {\bibinfo
  {title} {{New binary pulsar constraints on Einstein-\ae{}ther theory after
  GW170817}},\ }\href {https://doi.org/10.1088/1361-6382/ac1a69} {\bibfield
  {journal} {\bibinfo  {journal} {Class. Quant. Grav.}\ }\textbf {\bibinfo
  {volume} {38}},\ \bibinfo {pages} {195003} (\bibinfo {year} {2021})},\
  \Eprint {https://arxiv.org/abs/2104.04596} {arXiv:2104.04596 [gr-qc]}
  \BibitemShut {NoStop}%
\bibitem [{\citenamefont {Franchini}\ \emph {et~al.}(2021)\citenamefont
  {Franchini}, \citenamefont {Herrero-Valea},\ and\ \citenamefont
  {Barausse}}]{Franchini:2021bpt}%
  \BibitemOpen
  \bibfield  {author} {\bibinfo {author} {\bibfnamefont {N.}~\bibnamefont
  {Franchini}}, \bibinfo {author} {\bibfnamefont {M.}~\bibnamefont
  {Herrero-Valea}},\ and\ \bibinfo {author} {\bibfnamefont {E.}~\bibnamefont
  {Barausse}},\ }\bibfield  {title} {\bibinfo {title} {{Relation between
  general relativity and a class of Ho\v{r}ava gravity theories}},\ }\href
  {https://doi.org/10.1103/PhysRevD.103.084012} {\bibfield  {journal} {\bibinfo
   {journal} {Phys. Rev. D}\ }\textbf {\bibinfo {volume} {103}},\ \bibinfo
  {pages} {084012} (\bibinfo {year} {2021})},\ \Eprint
  {https://arxiv.org/abs/2103.00929} {arXiv:2103.00929 [gr-qc]} \BibitemShut
  {NoStop}%
\bibitem [{\citenamefont {Adam}\ \emph {et~al.}(2021)\citenamefont {Adam},
  \citenamefont {Figueras}, \citenamefont {Jacobson},\ and\ \citenamefont
  {Wiseman}}]{Adam:2021vsk}%
  \BibitemOpen
  \bibfield  {author} {\bibinfo {author} {\bibfnamefont {A.}~\bibnamefont
  {Adam}}, \bibinfo {author} {\bibfnamefont {P.}~\bibnamefont {Figueras}},
  \bibinfo {author} {\bibfnamefont {T.}~\bibnamefont {Jacobson}},\ and\
  \bibinfo {author} {\bibfnamefont {T.}~\bibnamefont {Wiseman}},\ }\bibfield
  {title} {\bibinfo {title} {{Rotating black holes in Einstein-aether
  theory}},\ }\href@noop {} {\  (\bibinfo {year} {2021})},\ \Eprint
  {https://arxiv.org/abs/2108.00005} {arXiv:2108.00005 [gr-qc]} \BibitemShut
  {NoStop}%
\bibitem [{\citenamefont {Barvinsky}\ \emph {et~al.}(2016)\citenamefont
  {Barvinsky}, \citenamefont {Blas}, \citenamefont {Herrero-Valea},
  \citenamefont {Sibiryakov},\ and\ \citenamefont
  {Steinwachs}}]{Barvinsky:2015kil}%
  \BibitemOpen
  \bibfield  {author} {\bibinfo {author} {\bibfnamefont {A.~O.}\ \bibnamefont
  {Barvinsky}}, \bibinfo {author} {\bibfnamefont {D.}~\bibnamefont {Blas}},
  \bibinfo {author} {\bibfnamefont {M.}~\bibnamefont {Herrero-Valea}}, \bibinfo
  {author} {\bibfnamefont {S.~M.}\ \bibnamefont {Sibiryakov}},\ and\ \bibinfo
  {author} {\bibfnamefont {C.~F.}\ \bibnamefont {Steinwachs}},\ }\bibfield
  {title} {\bibinfo {title} {{Renormalization of Ho\v{r}ava gravity}},\ }\href
  {https://doi.org/10.1103/PhysRevD.93.064022} {\bibfield  {journal} {\bibinfo
  {journal} {Phys. Rev. D}\ }\textbf {\bibinfo {volume} {93}},\ \bibinfo
  {pages} {064022} (\bibinfo {year} {2016})},\ \Eprint
  {https://arxiv.org/abs/1512.02250} {arXiv:1512.02250 [hep-th]} \BibitemShut
  {NoStop}%
\bibitem [{\citenamefont {Barvinsky}\ \emph {et~al.}(2017)\citenamefont
  {Barvinsky}, \citenamefont {Blas}, \citenamefont {Herrero-Valea},
  \citenamefont {Sibiryakov},\ and\ \citenamefont
  {Steinwachs}}]{Barvinsky:2017kob}%
  \BibitemOpen
  \bibfield  {author} {\bibinfo {author} {\bibfnamefont {A.~O.}\ \bibnamefont
  {Barvinsky}}, \bibinfo {author} {\bibfnamefont {D.}~\bibnamefont {Blas}},
  \bibinfo {author} {\bibfnamefont {M.}~\bibnamefont {Herrero-Valea}}, \bibinfo
  {author} {\bibfnamefont {S.~M.}\ \bibnamefont {Sibiryakov}},\ and\ \bibinfo
  {author} {\bibfnamefont {C.~F.}\ \bibnamefont {Steinwachs}},\ }\bibfield
  {title} {\bibinfo {title} {{Ho\v{r}ava Gravity is Asymptotically Free in 2 +
  1 Dimensions}},\ }\href {https://doi.org/10.1103/PhysRevLett.119.211301}
  {\bibfield  {journal} {\bibinfo  {journal} {Phys. Rev. Lett.}\ }\textbf
  {\bibinfo {volume} {119}},\ \bibinfo {pages} {211301} (\bibinfo {year}
  {2017})},\ \Eprint {https://arxiv.org/abs/1706.06809} {arXiv:1706.06809
  [hep-th]} \BibitemShut {NoStop}%
\bibitem [{\citenamefont {Griffin}\ \emph {et~al.}(2017)\citenamefont
  {Griffin}, \citenamefont {Grosvenor}, \citenamefont {Melby-Thompson},\ and\
  \citenamefont {Yan}}]{Griffin:2017wvh}%
  \BibitemOpen
  \bibfield  {author} {\bibinfo {author} {\bibfnamefont {T.}~\bibnamefont
  {Griffin}}, \bibinfo {author} {\bibfnamefont {K.~T.}\ \bibnamefont
  {Grosvenor}}, \bibinfo {author} {\bibfnamefont {C.~M.}\ \bibnamefont
  {Melby-Thompson}},\ and\ \bibinfo {author} {\bibfnamefont {Z.}~\bibnamefont
  {Yan}},\ }\bibfield  {title} {\bibinfo {title} {{Quantization of Ho\v{r}ava
  gravity in 2+1 dimensions}},\ }\href
  {https://doi.org/10.1007/JHEP06(2017)004} {\bibfield  {journal} {\bibinfo
  {journal} {JHEP}\ }\textbf {\bibinfo {volume} {06}},\ \bibinfo {pages}
  {004}},\ \Eprint {https://arxiv.org/abs/1701.08173} {arXiv:1701.08173
  [hep-th]} \BibitemShut {NoStop}%
\bibitem [{\citenamefont {Barvinsky}\ \emph {et~al.}(2022)\citenamefont
  {Barvinsky}, \citenamefont {Kurov},\ and\ \citenamefont
  {Sibiryakov}}]{Barvinsky:2021ubv}%
  \BibitemOpen
  \bibfield  {author} {\bibinfo {author} {\bibfnamefont {A.~O.}\ \bibnamefont
  {Barvinsky}}, \bibinfo {author} {\bibfnamefont {A.~V.}\ \bibnamefont
  {Kurov}},\ and\ \bibinfo {author} {\bibfnamefont {S.~M.}\ \bibnamefont
  {Sibiryakov}},\ }\bibfield  {title} {\bibinfo {title} {Beta functions of (3+
  1)-dimensional projectable ho{\v{r}}ava gravity},\ }\href@noop {} {\bibfield
  {journal} {\bibinfo  {journal} {Physical Review D}\ }\textbf {\bibinfo
  {volume} {105}},\ \bibinfo {pages} {044009} (\bibinfo {year}
  {2022})}\BibitemShut {NoStop}%
\end{thebibliography}%

\end{document}